\documentclass[prl,floats,showpacs,twocolumn]{revtex4}
\usepackage{graphicx}

\begin{document}
\title{\textbf{Generation and Bistability of a Waveguide Nanoplasma Observed\\ by Enhanced Extreme-Ultraviolet Fluorescence}}

\author{Murat Sivis}
\author{Claus Ropers}
\email{cropers@gwdg.de}
\affiliation{IV. Physical Institute, University of G\"ottingen, 37077 G\"ottingen, Germany}

\begin{abstract} 
We present a study of the highly nonlinear optical excitation of noble gases in tapered hollow waveguides using few-femtosecond laser pulses. The local plasmonic field enhancement induces the generation of a nanometric plasma, resulting in incoherent extreme-ultraviolet fluorescence from optical transitions of neutral and ionized xenon, argon, and neon. Despite sufficient intensity in the waveguide, high-order harmonic generation is not observed. The fluorescent emission exhibits a strong bistability manifest as an intensity hysteresis, giving strong indications for multistep collisional excitations.
\end{abstract}

\maketitle

Recent years have shown significant progress in translating concepts from attosecond and strong-field science~\cite{Corkum2007} into the areas of ultrafast nano-optics and plasmonics. Numerous characteristic nonlinear optical phenomena, such as above-threshold and strong-field photoemission, have become accessible using local field enhancements in metal nanotips~\cite{Bormann2010,Schenk2010,Kruger2011,Herink2012,DJPark2012}, resonant optical antennas~\cite{Kim2008,Sivis2012,Dombi2013,Sivis2013}, nanoparticles~\cite{Zherebtsov2011}, rough surfaces~\cite{Racz2011}, and plasmonic waveguides~\cite{Park2011,Park2012}. These experimental studies are paralleled by increasing theoretical efforts of describing strong-field effects at surfaces and in optical near fields~\cite{Yalunin2011,Ciappina2012,Yalunin2013}, with a particular emphasis on nanostructure-enhanced high-order harmonic generation (HHG)~\cite{Husakou2011_PRA,Husakou2011_OE,Stebbings2011,Shaaran2012,Yavuz2012,Ciappina2012a,Shaaran2012a,Ciappina2013,Fetic2013,Yang2013,Yavuz2013}.
Currently, some debate exists about experimental reports of nanostructure-enhanced HHG~\cite{Kim2008,Park2011,Park2012}, following the demonstration of the predominance of incoherent extreme-ultraviolet (EUV) fluorescence caused by multiphoton and strong-field atomic excitation and ionization~\cite{Sivis2012,Sivis2013}. The essential physical issue revolves around the tendency of gaseous media in nanoscopic generation volumes to favor incoherent over coherent processes. Thus, although not all expectations on realizing novel forms of attosecond spectroscopy in compact devices may ultimately materialize experimentally, the discussion reopens the field and further highlights the specificities and alternative scalings exhibited by strong-field interactions in nanostructures. This will allow for entirely new perspectives and phenomena in near-field high-intensity atomic and molecular gas excitations.

In this Letter, we demonstrate the nanostructure-enhanced formation of a confined plasma within a gas-filled plasmonic waveguide using low-energy, few-femtosecond laser pulses. This nanoplasma exhibits a pronounced threshold behavior and bistability of the fluorescent EUV emission that depends on the gas species and pressure. Specifically, we investigate the optical excitation, ionization, and emission characteristics of xenon, argon and neon atoms and ions, finding a strong intensity hysteresis of the fluorescence signal. Moreover, the exclusive observation of fluorescent emission highlights physical limits of nanostructure-based HHG concepts. Our results illustrate the plasmonic enhancement of strong-field effects in gaseous media and provide insight into the mechanisms associated with such phenomena driven at high repetition rates and in nanoscale confinement.

\begin{figure}[t!]
\centering
  \includegraphics[width=0.9\columnwidth,clip]{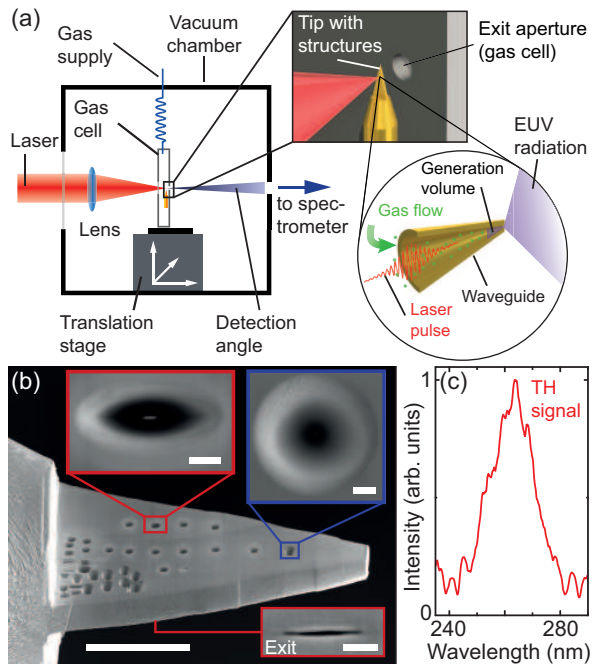}
  \caption[Schematics of the experimental setup/SEM images of the waveguide nanostructures/Third harmonic signal]{(a) Schematics of the experimental setup. The generated radiation in a solid angle of $\pm1.2^\circ$ is spectrally analyzed with a flat-field EUV spectrometer. (b) Scanning electron micrograph of the gold platform containing the tapered waveguides. Upper insets: Close-up views of the entrance apertures of an elliptical and a circular waveguide. Lower inset: Exit aperture of the elliptical waveguide. (Scale bars: overview 50~$\mu$m, top views 1~$\mu$m, exit aperture 200 nm.) (c) Third harmonic signal generated in a waveguide without gas exposure.}
  \label{sch:figure1}
\end{figure}

The setup of our experiment is depicted in Fig.~\ref{sch:figure1}(a). An electrochemically etched gold tip containing the waveguide structures [cf. Fig.~\ref{sch:figure1}(b)] is placed in a cell that can be purged with different gases. The structures are mounted on microtranslation stages (three translational and one rotational positioner) in a high vacuum chamber. Optical excitation with variable linear polarization is provided by 8-fs pulses (800~nm center wavelength) from a 78~MHz Ti:sapphire laser oscillator, focused to a 5~$\mu$m spot size (FWHM). Incident peak intensities up to about 1~TW/cm$^2$ are achieved. A 200~$\mu$m aperture in the gas cell allows for differential pumping of the detection chamber and the collection of the generated radiation within a solid angle of $\pm 1.2^{\circ}$. Spectral analysis of the collected emission is carried out with a home-built EUV flat-field spectrometer and a phosphor screen microchannel-plate detector. A more detailed description of the detection scheme is given elsewhere~\cite{Sivis2012,Sivis2013}.

Various waveguides with circular and elliptical cross sections are prepared on a 10~$\mu$m thick gold platform using focused ion beam etching [cf. Fig.~\ref{sch:figure1}(b)], and we have taken special care to produce smooth waveguide surfaces for minimal losses. For waveguides with elliptical entrance and exit apertures, geometrical dimensions are chosen that approximately correspond to those presented in Ref.~\cite{Park2011}, in which simulations indicated intensity enhancements of up to 350 near the waveguide end. As an example, Fig.~\ref{sch:figure1}(b) (top left inset, red) shows a waveguide with diameters of the minor  and major axes of the entrance (exit) apertures of 2.4~$\mu$m~(40~nm) and 4.8~$\mu$m~(440~nm), respectively.

Previously, optical second harmonic generation was used to characterize local fields at the exit aperture~\cite{Labardi2005} of hollow near-field probes \cite{Mihalcea1996}. Here, we utilize surface-enhanced third harmonic (TH) generation as a confirmation of field enhancement and efficient waveguide coupling~\cite{Tsang1996,Lippitz2005,Hanke2009,Hentschel2012,Sivis2013}. Figure~\ref{sch:figure1}(c) displays a TH spectrum obtained from an illuminated waveguide in the absence of gas. To ensure a proper alignment and detection of any possible coherent contribution in the EUV radiation, the TH signal was optimized by adjusting the lateral positioning of the structures and the angle of incidence.

\begin{figure}[t!]
\centering
  \includegraphics[width=0.85\columnwidth,clip]{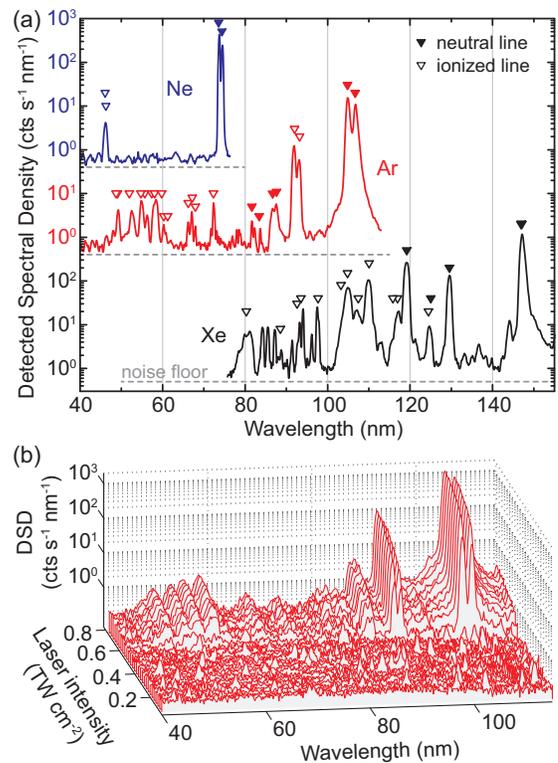}
  \caption[Nanostructure-enhanced EUV spectra]{(a) Nanostructure-enhanced EUV spectra (logarithmic scale) using xenon, argon and neon gas (200 mbar backing pressure) excited in the waveguide displayed in Fig.\ref{sch:figure1}(b) (inset, red frame). The argon and neon spectra are up shifted for better visibility. In all cases, dashed gray lines correspond to the same signal level (being the noise floor at 0.5~counts/(s~nm) in these measurements). Wavelength positions of triangles indicate the expected fluorescent transitions for neutral (filled) and singly ionized (open) atoms of the respective gas species~\cite{Sansonetti2005}. Providing for better resolution, the following portions of the spectra were recorded in the second grating diffraction order: Ne ($>$60~nm), Ar (65--84~nm), Xe (84--100~nm). (b) Intensity-dependent series of argon spectra recorded for increasing incident laser power.
  }
  \label{sch:figure2}
\end{figure}

\begin{figure}[t!]
\centering
  \includegraphics[width=0.7\columnwidth,clip]{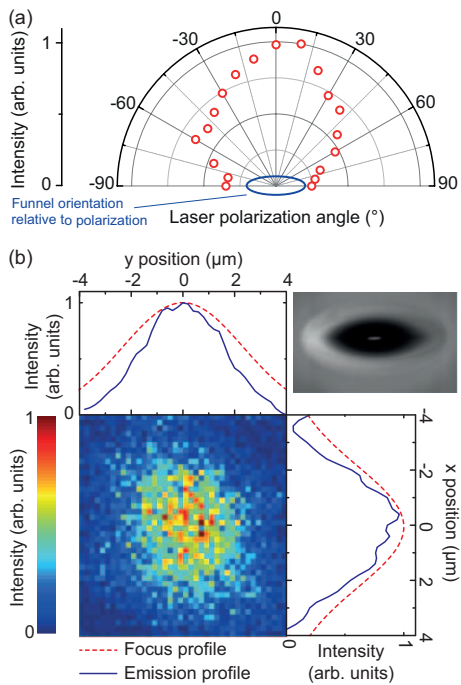}
 \caption[Polarization dependence and EUV emission map]{(a) Intensity of the 104.8-nm argon line as a function of incident linear laser polarization. Polarization parallel to the minor axis leads to maximum fluorescent emission.
(b) Emission map obtained for argon gas. The EUV fluorescence is recorded as the waveguide (cf. inset) is scanned relative to the focal spot. The averaged profiles in the $x$ and $y$ directions (solid blue) both show a significantly narrowed width compared with the focal laser profiles (dashed red), which is caused by spatial mode overlap in conjunction with the emission nonlinearity.
}
  \label{sch:figure3}
\end{figure}

In the following, we present our observations of the EUV radiation emitted from the illuminated structures exposed to atomic gases. Figure~\ref{sch:figure2}(a) shows EUV spectra measured using xenon, argon, and neon gases in the elliptical waveguide displayed in Fig.~\ref{sch:figure1}(b) (top left inset, red). The spectra exhibit numerous emission lines at the atomic and ionic transitions of the respective elements (indicated by triangles \cite{Sansonetti2005}) and thus present incoherent characteristic EUV fluorescence, in other words, atomic line emission. One of the essential features of the hollow waveguide structures, which has not yet been experimentally studied, is the theoretically proposed improvement in the field enhancement of an elliptical over a circular cross section~\cite{Park2011} and potential polarization dependencies. We have observed only minor EUV signals from waveguides with circular cross sections, such as the one displayed in Fig.~\ref{sch:figure1}(b) (top right inset, blue). Moreover, we have studied the dependence of the EUV emission on incident laser polarization [cf. polar diagram in Fig.~\ref{sch:figure3}(a)]. Maximum EUV intensity was found for a polarization parallel to the minor waveguide axis (corresponding to an angle of 0$^\circ$), which is consistent with an interpretation that the fundamental excitation for polarization parallel to the minor axis reaches deeper into the waveguide and thus creates higher overall field enhancement.

Despite numerous tests, accurate alignments [cf. Fig~\ref{sch:figure3}(b)], and recording more than 2000 individual EUV spectra using different waveguides, intensities, gases, and pressures, we have not found any indications for high harmonic generation in this spectral range, which appears to contradict the conclusions of previous reports~\cite{Park2011,Park2012}. For better comparison, let us briefly note the differences between our structures and those presented in Ref.~\cite{Park2011}. In principle, our use of gold instead of silver as waveguide material may cause higher absorptive losses~\cite{Barnes2003,Stockman2004,Maier2007}. However, since the typical propagation lengths of surface plamson polaritons in gold waveguides are still much larger than the length of the waveguides used~\cite{Bozhevolnyi2005,Maier2007}, we believe that structural quality and durability play a far more significant role than material selection for achieving stable strong-field conditions.
In this respect, using a bulk gold support allowing for smooth waveguide walls and high thermal conductivity has advantages over the silver-coated semiconductor taper previously used. Concerning the dimensions of the waveguide exit apertures [here several waveguides with minor (major) axes of typically about 50~nm~(400~nm) compared with 100~nm~(200~nm) in Ref.~\cite{Park2011}], we believe that the details of the aperture geometry are less critical for the overall volume in which strong-field conditions can be achieved, as long as substantial field compression in the waveguide is present. Instead, the exit aperture is expected to strongly influence the overall spectral transmittance and diffraction properties. Importantly, the field-enhanced volume in these structures is assumed to span at least several hundred nanometers before the end of the taper, i.e., multiple EUV wavelengths. Thus, even for incoherent fluorescent emission, wavelength-dependent divergence angles (previously interpreted as evidence for HHG \cite{Park2011}) are expected.

We note that the absence of HHG is not caused by insufficient local intensities, as is evident by the strong contributions from ionic transitions in our measured spectra. Substantial ionization is found even in neon with an ionization energy of 21.5~eV, exceeding the energy of 13 laser photons.  Furthermore, the observation of a predominance of incoherent emission is consistent with our previous studies on plasmonic bow-tie antennas, in which near-field strengths generally allowing for HHG were verified by utilizing the intensity-dependent atomic line emission spectral fingerprint~\cite{Sivis2013}. By applying this field gauging concept to the waveguide measurements, sufficient local peak intensities above several tens of TW/cm$^2$ are found. However, considering the field-enhanced volumes in both types of structures, severe limitations on high harmonic signal generation arise from the nanometric confinement of the source and the resulting insufficient coherent radiation buildup. Estimates on the relevant scalings are discussed in Refs.~\cite{Sivis2012,Sivis2013, Raschke2013}. Specifically, a quadratic scaling of the HHG yield with propagation length poses an important constraint in nanoscale geometries. Reference measurements using amplified laser pulses allow for an estimate of relative HHG to fluorescence signal levels [cf. Eq. (1) in Ref.~\cite{Sivis2013}]. In this comparison, diffraction of potential HHG also has to be considered. The total power of high harmonic radiation generated in the waveguide ($L_{nano}\approx500$~nm) is expected to be 2 orders of magnitude below that of fluorescence emitted into the $\pm 1.2^{\circ}$ detection angle. Diffraction of potential coherent radiation at the exit aperture further reduces the intensity in the forward direction by at least another 2 orders of magnitude. Thus, based on the signals presented in Fig.~\ref{sch:figure2}(a), we could expect to count less than one HHG photon per second and nanometer, which is below the presently detectable limits, and far below previously reported values. Supporting these considerations, an independent estimate for waveguide structures has projected the high harmonic power to be at least 8 orders of magnitude lower than that achieved in cavity measurements~\cite{Raschke2013,Cingoz2012}.

It seems clear that our findings present a serious setback for current approaches to realize compact nanoscale sources of high harmonic radiation for attosecond spectroscopy. Nonetheless, it is apparent that the strong-field regime is reached by means of field enhancement in these nanostructures. The long-term stability and reproducibility of the highly nonlinear EUV emission allows for further discovery in detailed investigations of the waveguide excitation and radiation mechanisms under such conditions.
\begin{figure}[b!]
\centering
  \includegraphics[width=0.85\columnwidth,clip]{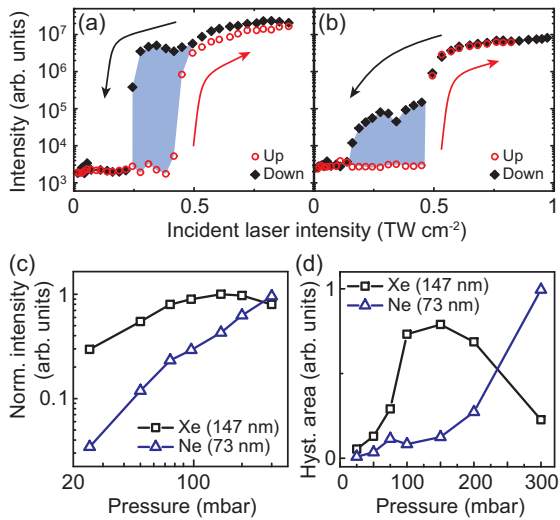}
   \caption[Intensity-dependent hysteresis and bistability of the EUV signal]{(a),(b) Intensity-dependent strength of the 104.8-nm argon line excited in two different waveguides (pressure: 200 mbar). A pronounced intensity hysteresis is observed. (c) Pressure-dependent intensities of two prominent xenon and neon lines. (d) Degree of bistability in xenon and neon as a function of gas pressure, measured as the normalized area enclosed by the hysteresis curves [cf. shaded areas in (a) and (b)].}
  \label{sch:figure4}
\end{figure}

To this end, we have carried out an analysis of the intensity and pressure dependencies of the emission for different gases. First, we find that as function of incident intensity---particularly for higher pressures---the fluorescent intensity exhibits a steep threshold behavior [displayed in Fig.~\ref{sch:figure2}(b)]. This indicates a density-dependent plasma formation in the waveguide that is most likely triggered by cascaded (for example collisional) excitation and ionization processes involving multiple excited atoms, ions, or electrons. Moreover, at higher pressures, we find a pronounced and reproducible intensity hysteresis, i.e., a situation in which the fluorescent signal at a given incident intensity depends on the excitation history of the system. Figures~\ref{sch:figure4}(a) and (b) display these intensity hysteresis curves for two different waveguide structures. Specifically, after having applied high intensity, substantial fluorescence can be maintained (upper branch) upon significantly decreasing the intensity to levels, at which hardly any fluorescence is detected when measured in conditions starting from low intensities (lower branch).
In atomic gases, bistable behavior is a characteristic feature of complex plasma dynamics, which can manifest itself as hysteresis in the plasma properties upon cycling various physical parameters. The familiar effect of maintaining a gas discharge for reduced voltages after ignition is just one example of such hystereses~\cite{Kortshagen1996,Turner1999,Ostrikov2000,Chen2007}.

The present observations constitute a new kind optical bistability \cite{Gibbs1985} within metal nanostructures~\cite{Wurtz2006,Shen2008,Wang2011}, which warrants further investigation. As expected for a bistable plasma effect, we find that it strongly depends on the gas species and pressure. In order to quantify the hysteresis, in Fig.~\ref{sch:figure4}(d), we evaluate the area enclosed by the two branches, normalized to the maximum fluorescent signal [Fig.~\ref{sch:figure4}(c)]. The prominence of the bistability is found to increase more rapidly with increasing pressures for xenon compared with neon. Moreover, whereas it decreases beyond 150 mbar pressure in xenon, neon displays increasing hysteresis throughout the measured pressure range. The earlier onset of hysteresis for xenon is understood by considering that it has both a lower ionization threshold and a larger atomic cross section for scattering events than neon. This results in reaching a critical excitation density for cascaded plasma formation at lower pressures.

A quantitative modeling of the plasma dynamics within the waveguide is outside of the scope of the present work, so that we give a qualitative description of the types of processes that may contribute to the observed bistability. Generally, the fact that a hysteresis is observed demonstrates that upon arrival of a laser pulse, there is residual excitation left within the waveguide from the previous excitation.  After the laser pulse period of approximately 12~ns, the state of the system will be composed of partially decayed populations of excited states in neutral atoms and ions, free electrons, as well as thermal excitation of the gas and waveguide walls. As thermal transport in the bulk waveguide is rather efficient, the likely most relevant forms of residual excitation are long-lived atomic and ionic states. The reduced ionization energy of such states then causes a higher ionization rate than without prior excitation. An excitation memory alone, however, is not sufficient for the development of a bistability. Instead, the bistable behavior is caused by a collective aspect of the plasma dynamics, which influences the global state of the system and depends in a highly nonlinear way on the density of excited atoms or electrons.

Whereas other potential plasma effects could contribute to the observed bistability, the nonlinear pressure dependence of the hysteresis and the fact that accelerated electrons from strong-field photoionization with energies up to several tens of electronvolts are present in significant density suggest that multistep collisional excitations may be the origin of the bistability, as in discharge plasmas~\cite{Turner1999}. Moreover, multiphoton and strong-field photoemisson from the waveguide walls~\cite{Kruger2011,Racz2011,Herink2012} will provide additional electrons capable of collisional excitation. We have found that upon moderate waveguide modification by using incident powers in excess of 200 mW, the steepness of the threshold slope could be influenced, potentially by introducing roughness and thus altering the contributions from surface-emitted electrons. A more detailed theoretical examination of the plasma bistability based on coupled rate equations~\cite{Vorobev1987,Turner1999} will be the subject of future work.\\

In conclusion, we have demonstrated the formation of a submicron noble-gas plasma within tapered waveguides by the observation of extreme-ultraviolet fluorescence. The results illustrate that such tapered plasmonic waveguides are not suitable for efficient nanostructure-enhanced HHG, although sufficient local intensities are achieved. We believe that coherent radiation buildup is limited by the nanoscopic generation length, which is orders of magnitude smaller than in other state-of-the-art HHG concepts~\cite{Cingoz2012,Li1989,Gibson2003,Gohle2005,Popmintchev2010}. Nonetheless, the observed strong EUV fluorescence and the exhibited optical bistability indicate novel aspects of strong-field physics in conjunction with nonlinear plasmonics. Further studies will elucidate the different roles of surface-mediated, collisional and direct excitation processes within the waveguides. We believe that the possibility to induce highly nonlinear processes in these devices may lead to applications in the areas of EUV near-field imaging, lithography and spectroscopy. The identified bistability of the EUV emission indicates cascaded excitation processes and presents a new link between nanostructure plasmonics and laser-induced surface plasma physics.

We thank M. Duwe, B. Abel, V. Radisch, B. Schr\"oder, and S. Sch\"afer for helpful discussions. Financial support by the Deutsche Forschungsgemeinschaft (DFG-ZUK 45/1 and SFB 755) is gratefully acknowledged.

\end{document}